\documentclass[aps,prl,amsmath,amssymb,reprint,superscriptaddress,twocolumn,showpacs]{revtex4-1}

 \usepackage{graphicx, subfigure}
 \usepackage{amsmath}
 \usepackage{amsfonts}
 \usepackage{amssymb}
 \usepackage{pstricks}
 \usepackage{psfrag}
 \usepackage{epsfig}
 \usepackage{changes}
\usepackage[ansinew]{inputenc}
\usepackage{mathptmx}
\usepackage{helvet}
\usepackage{textcomp}
\usepackage{ulem}
\usepackage{french}
\DeclareMathOperator{\csch}{csch}
\begin{document}


\title{
Singular Bifurcations : a Regularization Theory}


\author{Alexander Farutin}
\affiliation{Universit\'e Grenoble Alpes, CNRS, LIPhy, F-38000 Grenoble, France}
\author{Chaouqi Misbah}\email[]{chaouqi.misbah@univ-grenoble-alpes.fr}
\affiliation{Universit\'e Grenoble Alpes, CNRS, LIPhy, F-38000 Grenoble, France}


\date{\today}

\begin{abstract}
Several nonlinear and nonequilibrium driven  as well as active systems (e.g. microswimmers) show bifurcations from one state to another  (for example a transition from a non motile to motile state for microswimmers)  when some control parameter reaches a critical value. Bifurcation analysis  relies either on a regular perturbative expansion close to the critical point, or on a direct numerical simulation. While many systems exhibit a regular bifurcation  such as a pitchfork one, other systems undergo a singular bifurcation not falling in the classical nomenclature, in that the bifurcation normal form is not analytic.  We present a swimmer model which offers an exact solution showing a singular normal form, and serves as a guide for the general theory.
 We provide an adequate general regularization theory that allows us to handle properly the limit of singular bifurcations, and provide several explicit examples of normal forms of singular bifurcations.
  This study fills a longstanding gap in  bifurcations theory.\end{abstract}


\maketitle




\paragraph{Introduction.---}
Nonequilirium driven systems constitute a large branch of science which has been the subject of active research in the last decades \cite{Cross1994,Hoyle,Misbah,CrossBook}. Typical examples are B\'enard convection \cite{Goluskin}, Turing patterns \cite{Turing,Bourgine,Misbah}, crystal growth \cite{Kassner,Saito,Misbahreview} and so on.  By varying a control parameter (e.g. Rayleigh number in convection) the system exhibits a bifurcation from one state (e.g. quiescent fluid) into a new state (the fluid shows convection rolls) when a control parameter reaches a critical value. If $\mu$ designates the distance of the control parameter from the bifurcation point, the amplitude of the field of interest, say convection amplitude $A$, behaves as  $A\sim \pm \mu^{1/2}$, known as a pitchfork (classical) bifurcation.  In other situations, like driven fluids in a pipe,   a transition from a laminar to a non-laminar (turbument) flow takes place beyond a critical Reynolds number in the form of a saddle-node bifurcation \cite{Turbulence}.
This behavior is also generic  for many pattern-forming systems\cite{Cross1994,Hoyle,Misbah,CrossBook}. The dynamics of the amplitude $A$ (known as normal form) of these two bifurcations (pitchfork and saddle-node) read respectively 
\begin{equation}
\label{bifurcation}
\dot A=\mu A - A^3 ,\quad    \dot A =\mu -A^2
\end{equation} 
where dot designates time derivative.
Other types of bifurcations are also common, such as transcritical, subcritical\cite{Cross1994,Hoyle,Misbah,CrossBook} and so on. A hallmark of classical bifurcations theory is the regular (analytic) expansion in powers of $A$ in Eq.(\ref{bifurcation}). The same  holds also in catastrophe theory { \`a} la Ren\'e Thom \cite{Thom}.

More recently, active matter, a subject of great topicality, has  revealed several bifurcations from a non-motile state to a motile one when activity reaches a critical value \cite{izri2014self,MLB13,rednikov1994drop,Jin2017,Hu2019,morozov2019nonlinear,Izzet2020,Hokmabad2021,Chen2021}. In its simplest version this consists of a particle emitting/absorbing a solute which diffuses and is advected  in the suspending fluid. If the emission/absorption rate exceeds a critical value the particle transits from a non-motile to a motile state. The amplitude of the swimming velocity is found to behave (for infinite system size)  \cite{rednikov1994drop,Morozov2019JFM,Saha2021} as $\lvert A \rvert \sim \mu $ (or $A\sim \pm \mu$, $\mu>0$). This is a singular bifurcation behavior as encoded in $\lvert A\rvert $. 
In a marked contrast with the classical picture represented by (\ref{bifurcation}) the corresponding normal form reads
\begin{equation}
\label{bifurcations}
\dot A=\mu A- A\lvert A\rvert 
\end{equation} 
This means that  the  regular amplitude expansion ceases to be valid, as manifested by the non analytical term $\lvert A\rvert$. Numerical simulations \cite{Hu2019} of this system are, in contrast, consistent with a classical pitchfork bifurcation,   $A\sim \pm \mu^{1/2}$. We will see that this is due to finite size in numerical simulations. 

Examples of singular nature have been also encountered in crystal growth.  
 It has been shown that the usual perturbative scheme in terms of the crystal surface deformation amplitude is not legitimate \cite{OPL}. 
 Besides  these examples, the emergence of singular bifurcations is  likely to be abundant, and has been probably overlooked in many numerical simulations (see also conclusion). 
The purpose of this Letter is to fill this gap.
 We will show how to handle singular bifurcations from the usual commonly used  regular perturbative scheme.
We will first illustrate  the theory on an explicit example of microswimmer for which an exact analytical solution is obtained.  We then present  a systematic  method on how  to properly  treat singular bifurcations.

\paragraph{Theory---} 
It is instructive to begin with an explicit model revealing a singular bifurcation. We first introduce the full model, before considering a simplified version which can be handled fully analytically.
The model consists \cite{MLB13} of a rigid  particle (taken  to be a   sphere with radius $a$), which emits/absorbs a solute that diffuses and is advected by the flow. 
The advection-diffusion equations read  as   
\begin{equation}\label{Eq:advection-diffusion}
\frac{\partial c}{\partial t} + \mathbf{u}\cdot\nabla c = D\Delta c,\;\; 
\end{equation}
where $c$ is the solute concentration, $D$ is the diffusion constant,  $\mathbf{u}$  and $p$ are the velocity and pressure  fields, obeying Stokes equations.  
The associated boundary conditions of surface activity and the swimming speed (which will be taken to be along the $z$-direction)  are
\begin{equation}\label{BC}
D\frac{\partial c}{\partial r}(1,\theta,t) = -\mathcal{A},\;\;\;  {V_0} =-{{ {\cal M}\over a} }  \int _{-1}^1 \mu c(r=a,\mu ,t) d\mu \end{equation}
with $\mu=\cos(\theta)$, where $\theta$ is the azimuthal  angle in spherical coordinates, $a$ is the particle radius ${\mathcal A}$  is the emission rate ($\mathcal{A}>0$: emission, $\mathcal{A}<0$: adsorption),  $\mathcal{M}$ is a mobility factor (which can be either positive or negative); see \cite{MLB13} for more details. 
This model has been studied  numerically \cite{MLB13,morozov2019nonlinear,Hu2019}, coming to the conclusion that for $P_e$ (with   $Pe = |\mathcal{A}\mathcal{M}|a/D^2$) sufficiently small the only solution is the non-moving state of the particle, with a concentration field which is symmetric around the particle. When $Pe$ exceeds a critical value it is shown that the concentration field loses its spherical symmetry and a concentration comet develops, resulting in a motion of the particle with a constant velocity $V_0$. It is found numerically \cite{Hu2019} that 
$V_0$ is well represented by  $V_0\sim \sqrt{Pe-Pe_1}$, where $Pe_1$ is the critical value of $Pe$ at which the transition from  a non motile to a motile state occurs.  The determination of the critical condition has also been analyzed by linear stability analysis \cite{MLB13,morozov2019nonlinear,Hu2019}. Analytical asymptotic perturbative studies\cite{rednikov1994drop,Morozov2019JFM,Saha2021} (for an infinite system size) revealed that the velocity of the swimmer follows in fact the following singular behavior $ \lvert { V}_0\rvert  \sim  ( {Pe- Pe_1})$. 
\paragraph{Exactly solvable  model--}
The main simplification adopted here is to disregard the fluid, in that the variable $\mathbf u$ is ignored in what follows. A justification of this is the fact that the singular behavior    is associated with the concentration field  at long distance\cite{rednikov1994drop,Morozov2019JFM,Saha2021}, while the velocity field vanishes at infinite distance from the swimmer.  We consider a particle  moving at constant speed $\mathbf {V}_0$. A further simplification is that we assume that the particle size is  small in comparison to  length scales of interest. The only  length scale is given by $D/V_0$, so our assumption corresponds to assuming $a\ll D/V_0$. Under this assumption the particle can be taken as a quasi-material point.
 With these assumptions the corresponding simplified model reads (in the laboratory frame )
\begin{equation}\label{diff2}
\frac{\partial c}{\partial t} -  D\Delta c = S \delta (\mathbf{r}-\mathbf{V}_0 t)
\end{equation}
where $S$ is the emission rate  (related to $\mathcal{A}$, by ${\cal A} = S/(4\pi a^2)$) 

Using the diffusion propagator the solution is given by
\begin{equation}\label{diff3}
c(\mathbf {r},t)  = \int_0 ^\infty d\tau  {S \over (4\pi D\tau)^{3/2} }\exp- { \left \{ {(\mathbf{r}+\mathbf{V}_0\tau - \mathbf{V}_0t)^2\over  4D\tau} \right\} } \; ,\end{equation}
Expression (\ref{diff3}) can be integrated to yield 
\begin{equation}\label{cintegr}
c( \mathbf{\tilde r})= {S \over 4\pi D} {\exp { \left \{ -{ \mathbf{\tilde r}\cdot \mathbf{V}_0 + \lvert V_0\rvert {\tilde r}  \over 2D}  \right\}   } \over r }\end{equation}
with $\tilde {\bf r}= \mathbf{r}-\mathbf{V}_0 t$ (the coordinate in the frame attached to the particle).  Along $z$, it is clear that the concentration decays exponentially with distance ahead of the particle, while it decays only algebraically 	at the rear ($c$ has front-back symmetry). Indeed, the emitted solute is advected (by swimming speed) backwards, enriching the rear zone, whereas ahead of the particle only diffusion can be effective. 

Using  (\ref{BC}), only the first spherical harmonics enters the expression of velocity, and we obtain  $ {V}_0=- M c_1/ (a\sqrt{3\pi})$, $c_1$ being the first harmonic amplitude, obtained  
by projection of (\ref{cintegra}) on that  harmonic, so that the velocity satisfies \begin{equation}\label{cintegra}, 
\bar V_0=  4 Pe \; e^  { -\lvert \bar V_0 \rvert/2} \left[  { { \bar V_0\cosh( \bar V_0/2 )-2\sinh( \bar V_0/2 )\over \bar V_0^2}  }\right]  , \;\; \bar V_0\equiv {a   {V}_0 \over D }\end{equation}
 Expanding for small $\bar V_0$
 we obtain

\begin{equation}\label{V0p}
 \bar {V}_0= {Pe\over Pe_1}   \bar V_0 \left (1- 2 \lvert \bar V_0\rvert  \right ) 
\end{equation}
where ${\cal A} = S/(4\pi a^2)$, and $Pe_1=3$, is  the critical P\'eclet number. In the full model $Pe_1=4$ \cite{MLB13}. Including hydrodynamics close to particle surface we can capture analytically this result \cite{SM}.
 The result (\ref{V0p})  has been also obtained thanks to a singular perturbative scheme \cite{rednikov1994drop,Morozov2019JFM,Saha2021}.
We  see from (\ref{V0p}) that  $\bar V_0=0$ always exists. When $Pe>Pe_1$, there exists another solution  given by 
\begin{equation}\label{V0pp}
\lvert {\bar V}_0\rvert \simeq  {1\over 6}  ( {Pe- Pe_1}   )
\end{equation}

Expression (\ref{V0pp})  corresponds to a pitchfork bifurcation (and not trancritical \cite{Morozov2019JFM}) where the $\bar V_0=0$ solution becomes unstable in favor of two symmetric solutions, $ {\bar V}_0 \sim  \pm ( {Pe- Pe_1})$. This is, however,  an atypical behavior of a pitchfork solution, and is traced back to the infinite system size (as seen below). We refer to this bifurcation as {\it singular pitchfork bifurcation}. The term 'singular' refers to the non analytic nature $\lvert {\bar V}_0\rvert$.

Finite size regularizes the bifurcation and turns the singular bifurcation into a classical pitchfork bifurcation (see \cite{SM}). 
 Another way to regularize the model is via a consupmtion of solute in the bulk. In that case we modify Eq.(\ref{diff2}) by adding $\beta c$ on the left hide side,
where $\beta$ is the consumption rate. We have in mind the possibility that the emitted solute reacts in the bulk and is consumed by another reaction, giving rise to some secondary product.   The solution for $c$ becomes

\begin{equation}\label{cintegrp}
c( \mathbf{\tilde r})= {Sa \over 4\pi D} {\exp { \left \{ -{ (\mathbf{ \tilde r}/a)\cdot \mathbf{\bar V}_0 + \sqrt{\bar V_0^2+\epsilon^2}  \lvert { \tilde  r/a} \rvert \over 2}  \right\}   } \over (r/a) }, \end{equation}
with ${\epsilon}^2= {4a^2\beta \over D}$. The equation for $\bar V_0$ becomes 
\begin{equation} \label{Vreg} \bar V_0=4{ {Pe}  }\;   e^  { -\sqrt{ \bar V_0^2+\epsilon ^2}/2} \left[  {  {\bar V_0 \cosh(\bar V_0/2 ) - 2\sinh( \bar V_0/2) \over   \bar V_0^2} }\right]   \end{equation}

For  $\epsilon=0$ we recover the singular bifurcation solution, and for  $\epsilon\ne 0$ we obtain a regular pitchfork bifurcation. Expansion for small $\bar  V_0$ provides
\begin{equation}\label{expreg}
\bar V_0=Pe \bar V_0 e^{-\epsilon/2} \left \{ {1\over 3 } + {\epsilon -10\over 120 \epsilon} \bar V_0^2 + O(\bar V_0^4)\right \}
\end{equation}
Besides the trivial solution $\bar V_0=0$, we have $\bar V_0\sim \pm (Pe-Pe_1)^{1/2}$ which is a classical pitchfork bifurcation, with $Pe_1=3$. Consumption has turned the singular bifurcation into a regular bifurcation.
\paragraph  {Regularization theory--} 
The expression of type (\ref{expreg}) is the one that one would usually obtain by an analytical expansion in $\bar V_0$ in the absence of an exact solution. By trying to compare it to the exact solution  (\ref{Vreg}) in the vicinity of bifurcation where $\bar V_0$ is small  (Fig.\ref{regsing}) one realizes that the smaller $\epsilon$ is the worse the approximation (\ref{expreg})  is, and a fortiori this expression can in no way account for the singular limit $\epsilon=0$, a limit where the coefficients of  the  series (\ref{expreg}) diverge. One could then be tempted to say that  (\ref{expreg})  is of little practical interest for small $\epsilon$. However, and this is the main point, we will be able, in a way that may seem a little surprising, to extract from  analysis of a regular expansion (\ref{expreg})  the singular behavior $\lvert \bar V_0 \rvert $ (for $\epsilon\rightarrow 0$) dictated by   the exact calculation  (\ref{Vreg}), without any a priori knowledge on an exact solution. Moreover, we will  regularize the  expression (\ref{expreg}) in such a way that it represents correctly the exact behavior when $\epsilon$ is nonzero but small.

The crux of our theory is the observation that the singular behavior in the above model is due to the existence of a singular point in the complex plane, namely $V_0=i \epsilon$, arising from $\sqrt{\bar V_0^2+\epsilon^2}$ in (\ref{Vreg}). This model will serve as a precious guide, but the theory can be made general. We assume that the trivial solution always exist ($\bar V_0$ in the above model), so that the search for nontrivial solutions amounts to setting in  (\ref{Vreg}) the r.h.s. divided by $\bar V_0$ (to be denoted below as $f(\bar V_0,\epsilon)$) equal to unity. We focus on the behavior of $f(\bar V_0,\epsilon)$.
We use below the  notation  $f(x,\epsilon)$ to present the general theory. Suppose, without restriction,  that singularity is located on the imaginary axis at $x=i\epsilon$.
We propose the following transformation 
\begin{equation}  \label{svar}
\epsilon=x_0(1-s), \quad x^2=x_0^2(2s-s^2)
\end{equation} 
with $x_0$ a real positive number. Thanks  to this transformation $x^2+\epsilon^2=x_0^2$ remains constant. $s$ is a parametrization, and the singular limit corresponds to $s=1$.The above transformation means that  instead of taking the singular  limit $\epsilon \rightarrow 0$ at given $x$, we move in the plane $(x,\epsilon)$ along the circle of radius $x_0$. This transformation renders the expansion in terms of $s$ regular since $x^2+\epsilon^2$ is constant along the circle. Another way to appreciate our choice is  that the singularity in the original coordinate, $x^2=-\epsilon^2$, reads $x_0^2(1-(s-1)^2)=-x_0^2(1-s)^2$ which has no solution meaning that in terms of  $s$-variable the original singularity   has been  moved to infinity. This guarantees absolute convergence of series in term of $s$. The procedure consists now in substituting in the regular expansion 
\begin{equation}
\label{taylor}
 f(x,\epsilon) =\sum _{k=0}^\infty a_{k}(\epsilon)  x^{2k} 
  \end{equation} 
 $x$ and $\epsilon$ as functions of $s$ and $x_0$ (Eq.(\ref{svar})) and expand in a Taylor series in terms of $s$ as  
\begin{equation}
\label{sum}
 f(x(s),\epsilon (s)) =\sum_{k=0}^\infty a_k\left [ x_0(1-s)\right ] (2s-s^2)^{2k}  x_0^{2k}=\sum _k b_{k}(x_0) s^{k} 
  \end{equation} 
  The relation between $b_k$ and $a_k$ is easily deduced (see \cite{SM}). Close to the bifurcation  point   $x_0$ is small, so we will   retain only $b_0$, $b_1$ and $b_2$. Let us illustrate the study on the phoretic system. Taylor expansion of (\ref{Vreg}) to order $x^4$  (in the form (\ref{taylor})) yields
  \begin{eqnarray}  a_0(\varepsilon)&&={e^{-\varepsilon/2}\over 3}, \quad a_1(\varepsilon)=e^{-\varepsilon/2}{ 1-10/\varepsilon \over120}, \nonumber
 \\a_2(\varepsilon)&&=\frac{\varepsilon^3-28\varepsilon^2+140\varepsilon+280}{13440\varepsilon^3}e^{-\varepsilon/2}.   
  \end{eqnarray}
  from which $b_k's$ are determined and $f(\bar V_0,\varepsilon)$ reads  
  \begin{equation} \label{exp}   f(\bar V_0,0)=Pe  [1/3-|\bar V_0|/6+\bar V_0^2/20+O(\bar V_0^3)]  \end{equation}    
  A remarkable feature is that  due to our regularization theory we are able to extract, by using the traditional analytical expansion  (\ref{taylor}), the singular behavior exhibiting the absolute value $\lvert \bar V_0 \rvert$. Referring to the exact result obtained in the limit $\epsilon=0$ (Eq. \ref{cintegra})), we can check that to leading order in $\bar V_0$ we obtain exactly the result 
  (\ref{exp}) (recall we omit the trivial solution $\bar V_0=0$).This shows the consistency of  the  theory.  Another virtue of the theory is that it allows to transform the expansion (\ref{expreg}), which has a small radius of convergence of order $\epsilon$, into a form having a wider radius of convergence by applying the method above (used for $\epsilon\rightarrow 0$) for a finite $\epsilon =\epsilon^*$ with a corresponding value $x^*$. For that purpose we use  the substitutions $x_0= \sqrt{x{^*}^2+\varepsilon {^*}^2}$ and $s=1-\varepsilon ^*/    \sqrt{x{^*}^2+\varepsilon {^*}^2}$ in the second expression of (\ref{sum}). To leading order in $s$ we get $f=a_0(s=0)+ (2a_1(s=0)x_0^2-a_0'(s=0) x_0) s+O(s^2)$, where prime designates derivative with respect to argument. As an illustration for the phoretic model the function $f$ takes now the form 
   \begin{equation} \label{epsilons0}
 f=Pe{e^{{-\sqrt{\bar V_0^2+\varepsilon^2} / 2}}\over 3} \left [ 1 +    {\bar V_0  ^2+\varepsilon ^2 \over 20 } - {\varepsilon  \sqrt{\bar V_0 ^2+\varepsilon ^2} \over 20} \right ] \end{equation}
      instead of  (\ref{expreg}). We have now omitted the stars for simplicity. It can be checked that this expression reduces to (\ref{expreg}) after expansion in $\bar V_0$ to order 2.  Figure \ref{regsing}  summarizes the results. Use of expansion (\ref{expreg}) --dotted linesin Fig.\ref{regsing}--  fails to capture properly the  bifurcation from obtained from the exact result (Eq. (\ref{Vreg}), represented by  solid lines in Fig. \ref{regsing}), and tis becomes worst   as $\varepsilon $ goes to zero. In contrast  (\ref{epsilons0}) --dashed lines in Fig.\ref{regsing}-- impressively captures the exact result (Eq. (\ref{Vreg}), solid lines in Fig. \ref{regsing}). 
      The regularization theory does not only account properly for the singular limit ($\epsilon=0$; Eq.(\ref{exp})) but also it offers a precious way to approach this limit (Eq.\ref{epsilons0}). 
\begin{figure}[h]
\includegraphics[width = 0.9\columnwidth]{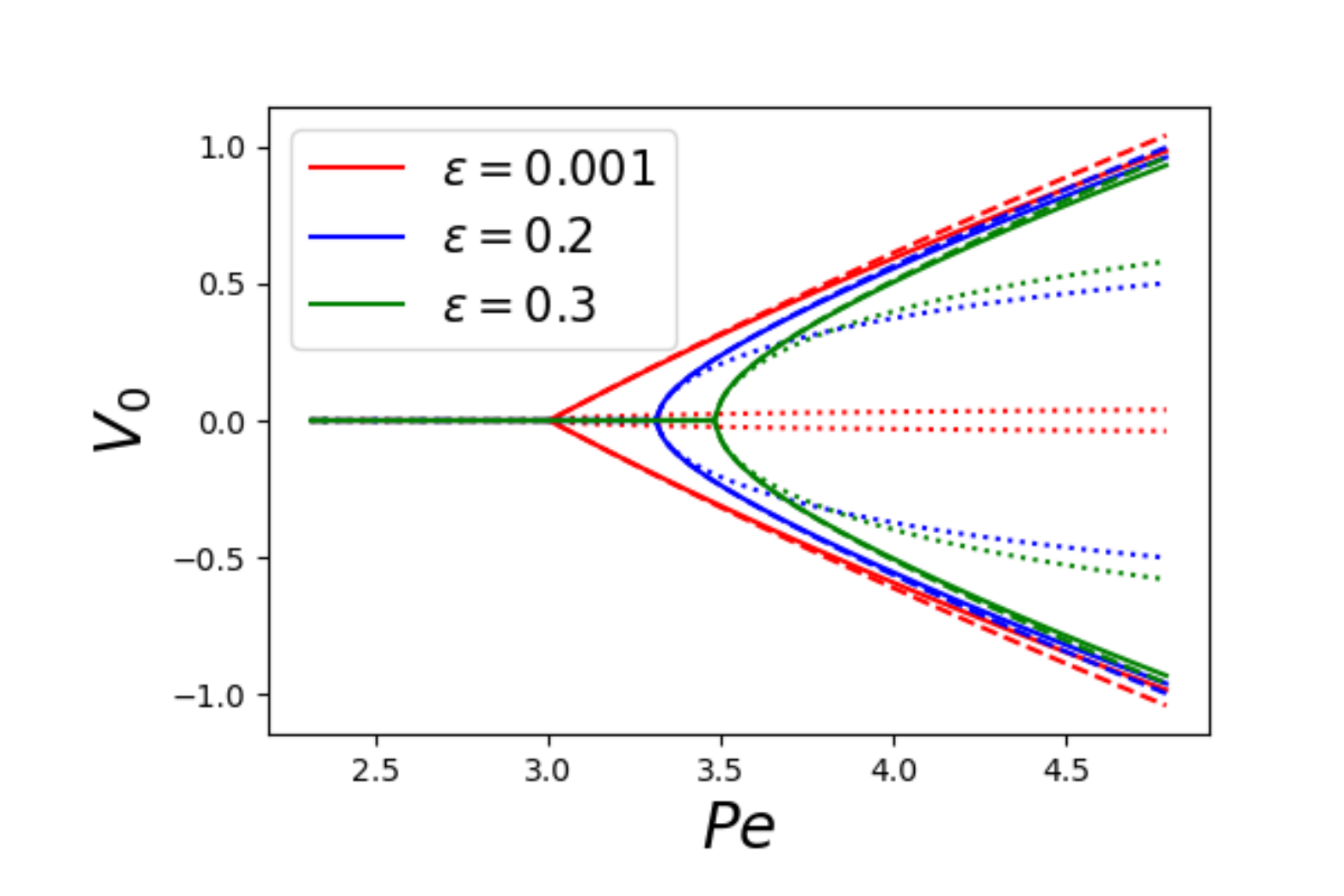}
\caption{Bifurcation diagram. solid line: exact result (\ref{Vreg}). Dotted  lines represent classically expanded solution Eq.(\ref{expreg}). Dashed lines represent regularized solution Eq.(\ref{epsilons0}). These dashed lines almost coincide  with solid lines (exact solution), despite that only leading order in Eq.(\ref{epsilons0}) is retained. \label{regsing}}
\end{figure}           

    Generally, in nonlinear systems an exact solution is the exception. The traditional way is then to expand the model equations in power series in an amplitude (denoted here as $x$) to obtain the final result in the form (\ref{taylor}). The present study shows that  wee can extract from  the traditional expansion thee results   (\ref{taylor}) (\ref{exp}) and (\ref{epsilons0}),  the correct singular behavior and the appropriate regularized form when $\epsilon$ is small but finite. This highlights the generality of the method and its application to various  nonlinear systems with a hidden singularity.
    
    Let us finally briefly classify  singular bifurcations  on the basis of the behavior of the general traditional expansion (\ref{taylor}). Suppose that the singularity is due to the presence of terms of the form $(x^2+\epsilon^2)^{\alpha}$ where $\alpha$ is real non integer positive number such that $\alpha<1$. Following the  general procedure presented above, we straightforwardly obtain to leading order
          \begin{equation} \label{epsilons1}
 f(x_0,0)=\beta - \lvert x_0 \rvert^{2\alpha}  
  \end{equation}  
  where $\beta$ is a real number, and where we have rescaled $x_0$ so that the coefficient in front of the singular term can be set to unity. If $\alpha>1$     the first dominant term is $x_0^2$ and to leading order the expansion is regular. Note that we have assumed the first nonlinear term to saturate the linear growth, this is why we set its coefficient to be negative. In the opposite case higher order terms (such as $x_0^2$) must be taken into account). This question is beyond our scope here.  In terms of a dynamical system, and by remembering that we assume $x_0=0$ to exist always as a  solution, the corresponding normal form is
       \begin{equation} \label{epsilons2}
 \dot A=\mu A - A\lvert A \rvert^{2\alpha}  
  \end{equation}   
    with $\mu =1-\beta$. Equation (\ref{epsilons2}) constitutes the generic normal form  for singular bifurcation. We used here the notation $A$, as often adopted in bifurcation theory. The nontrivial fixed point behaves as $A\sim  \pm  \mu^{1/(2\alpha)}$. The bifurcation structure is qualitatively different depending on whether $\alpha >1/2$ or $\alpha<1/2$. In the first case the bifurcation diagram is similar to a pitchfork bifurcation with infinite slope at $\mu=0$, whereas in the second case the slope vanishes for $\mu =0$. $\alpha=1/2$ is a special case with finite slope. Finally for $\alpha<0$ the normal form is 
      \begin{equation} \label{epsilons3}
 \dot A=\mu A +A\lvert A \rvert^{2\alpha}  
  \end{equation}     
  We adopted the positive sign in front of the nonlinear term to guarantee a stable branch for $A\ne 0$. Note that this does not affect the bifurcation diagram topology. The nontrivial fixed point is given $A\sim  \pm  (-\mu)^{1/(2\alpha)}$.
    Figure \ref{bifsing} summarizes the results. We note four different singular bifurcations (in blue  in Fig.\ref{bifsing}) corresponding to (i) $\alpha>1/2$, (ii) $\alpha<1/2$, (iii) $\alpha=1$, (iv) $\alpha<0$. We refer to these four singular bifurcations as (i) fold, (ii) cusp, (iii) angular and (iv) unbounded.
     When these bifurcations are regularized, they all fall into a pitchfork bifurcation (Figure \ref{bifsing}). We may refer to the above bifurcations as {\it singular pitchfork bifurcations} as well, albeit the singular limits have different behaviors. It must be noted that the above classification does not exhaust by far all kinds of singularities. For example, the 2D phoretic model provides an example of transcendental singularity where the velocity behaves as $\bar V_0\sim e^{-1/Pe}$ \cite{SM}. 
  \begin{figure}[h]
\centering
\includegraphics[width = 0.45\columnwidth]{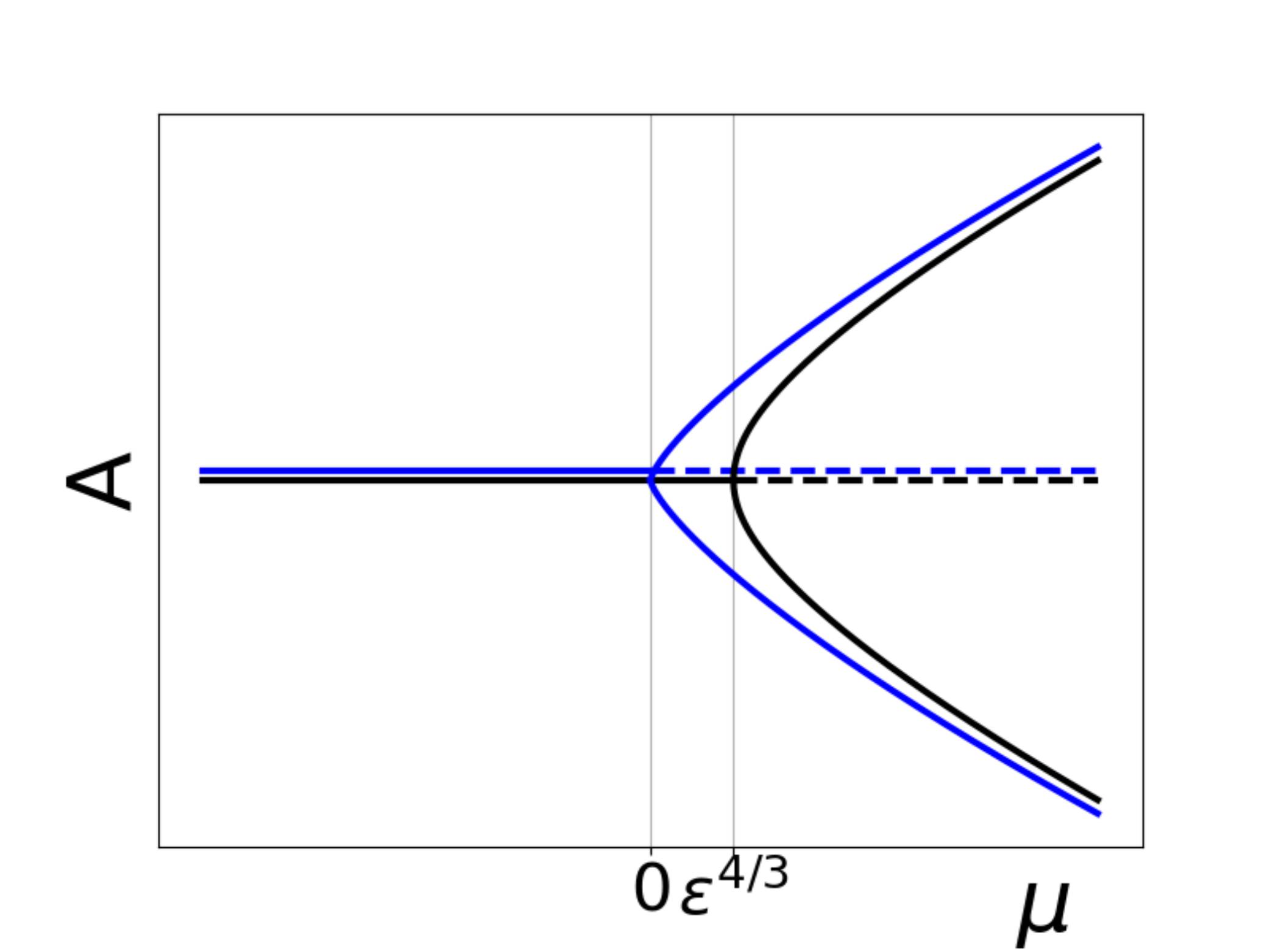}
\includegraphics[width=0.45\columnwidth]{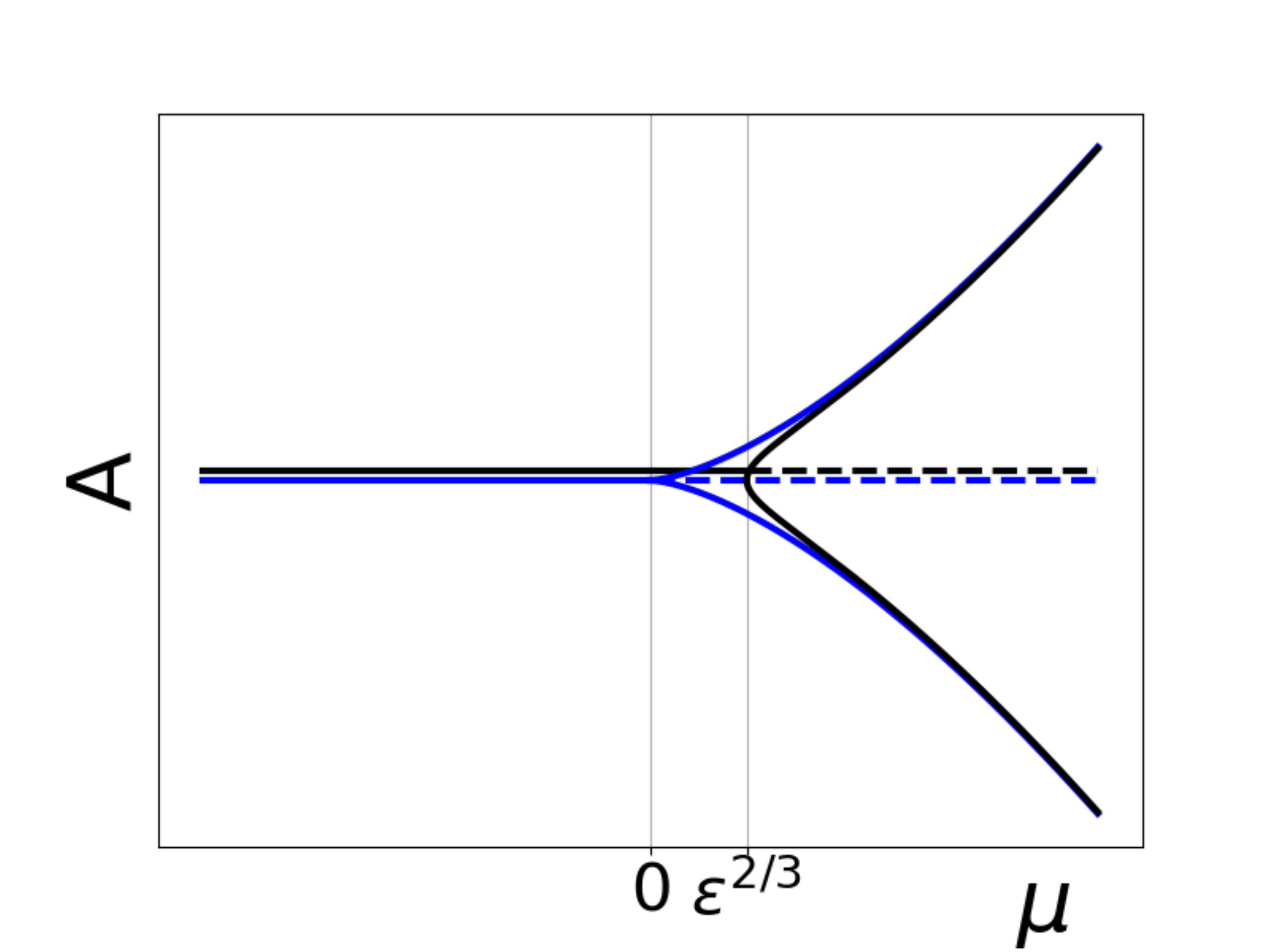}
\includegraphics[width=0.45\columnwidth]{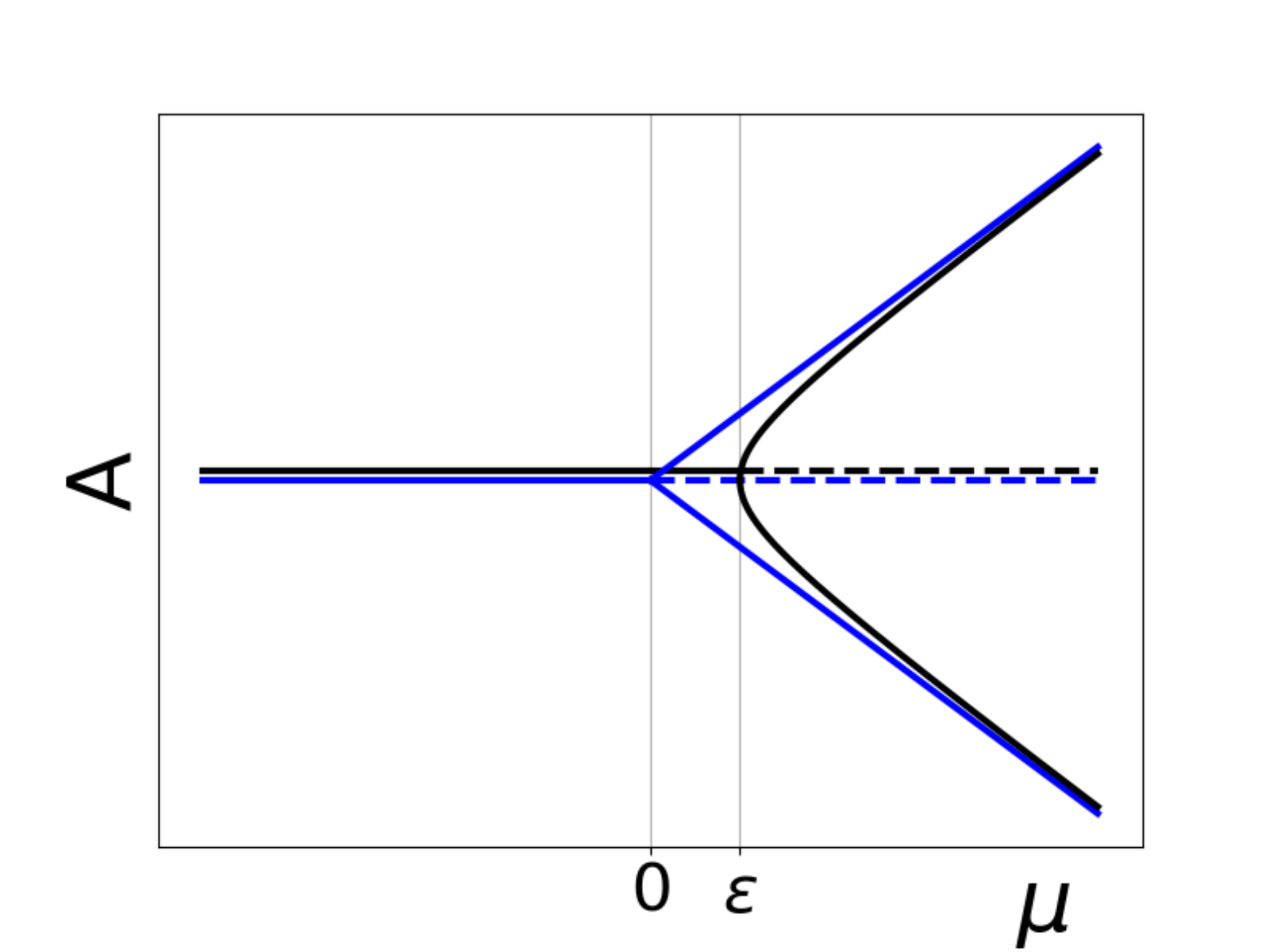}
\includegraphics[width=0.45\columnwidth]{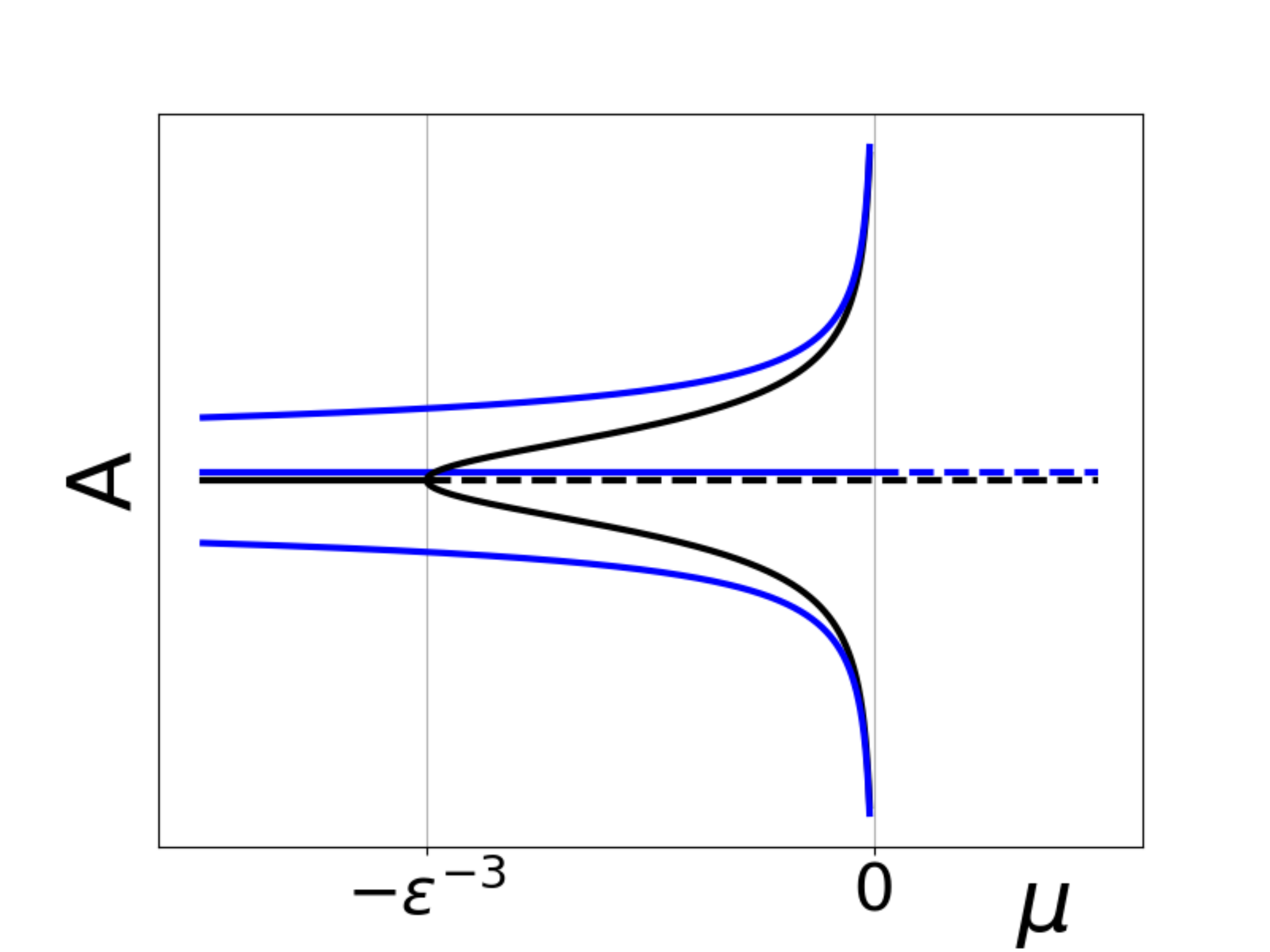}
\caption{Diagram for the four different singular bifurcations (in blue) and for their regularized form (in black), with $\alpha=2/3$, $\alpha=1/3$, $\alpha=1/2$ and $\alpha=-3/2$. Solid lines refer to stable solutions and dashed lines to unstable ones.\label{bifsing}}
\end{figure}   

Some important remarks are in order. We have assumed that the singularity of $f$ lies on the imaginary axis, $x=i\epsilon$. 
{ Note, however, that it can happen (as in the phoretic model with finite size; see \cite{SM}) that there  exists an infinite countable set  of singularities on the imaginary axis. It is also not excluded that there may be systems with several  singularities scattered in the complex plane, for which no general theory is at present   available}. However, in \cite{SM} we provide a condition of validity of the theory even when singularity does not lie on the imaginary axis.      

\paragraph {Conclusion--}
We have provided a framework to deal with singular bifurcations. The few concrete examples mentioned in the introduction are far from having exhausted all cases where singular bifurcations manifest themeselves.
 Suzade et al \cite{Sauzade} analyzed  the speed of the Taylor swimmer sheet in perturbation theory as a function amplitude of the swimmer deformation by including up to 1000 terms in the series expansion. They found that the series diverges beyond an amplitude of deformation (which is moderate). This is  symptomatic of a hidden singularity in the model.
{ In another problem, that of vesicles (a simple model of red blood cells) in a flow  \cite{Aouane,FarutinLaw}, the perturbative schemes for vesicle dynamics (in power series of excess area from a sphere) has a small range of applicability even when including higher and higher order terms in the series expansion. This is indicative of potential singularity in complex plane.} It is hoped that this study serves as a general framework to analyze singular bifurcations.

We thank CNES (Centre National d'Etudes Spatiales) for financial support and for having access to data of microgravity, and the French-German university programme "Living Fluids" (Grant CFDA-Q1-14) for financial support. 

%
\pagebreak
\widetext
\begin{center}
\textbf{\large Supplemental Materials: Singular Bifurcations: a Regularization Theory}
\end{center}
\setcounter{equation}{0}
\setcounter{figure}{0}
\setcounter{table}{0}
\setcounter{page}{1}
\makeatletter
\renewcommand{\theequation}{S\arabic{equation}}
\renewcommand{\thefigure}{S\arabic{figure}}
\renewcommand{\bibnumfmt}[1]{[S#1]}
\renewcommand{\citenumfont}[1]{S#1}
\title{
Supplementary Information: Singular Bifurcations : a Regularization Theory}




\date{\today}

We provide here the regularization solution for the phoretic model for finite size in 3D. We also present  the singular behavior in 2D, which is quite distinct from that in  3D. More details about the results discussed in the main text are also presented.


\maketitle

\section{Effect of hydrodynamics on critical condition}
The goal of this section is to introduce the corrections into the exactly solvable model in order to account for the finite size of the particle.
These corrections are evaluated for small propulsion velocity and provide quantitatively correct value of the critical Peclet number.
There are two finite-size effects that are neglected in the main model: First, the near-field flow disturbance due to a translating spherical particle is neglected, and second, the particle emission is represented by a point source, while the finite-size particle should be represented by a homogeneous distribution of sources along the particle surface.
Both of these two effects are essential for quantitative evaluation of the concentration field close to the critical point.

This problem is solved in the reference frame comoving with the particle.
The concentration evolution equation is then written as
\begin{equation}
\label{comoving}
\dot c(\boldsymbol r)+\boldsymbol\nabla\boldsymbol\cdot(\boldsymbol u(\boldsymbol r) c(\boldsymbol r))=D\nabla^2 c(\boldsymbol r)+A(\boldsymbol r),
\end{equation}
where $\boldsymbol u(\boldsymbol r)$ is the fluid velocity relative to the particle, $A(\boldsymbol r)$ represents a distribution of sources and source dipoles on the particle surface which accounts for the concentration emission or consumption, and $\boldsymbol r$ is the position vector relative to the particle center.
It is known that the velocity field in the comoving frame can be written as
\begin{equation}
\label{flow}
\boldsymbol u(\boldsymbol r)=-\boldsymbol V_0 +\frac{a^3}{2r^3}\left[3\frac{\boldsymbol r(\boldsymbol r\boldsymbol\cdot\boldsymbol V_0)}{r^2}-\boldsymbol V_0\right]
\end{equation}
for a rigid force-free spherical particle or radius $a$, moving with velocity $V_0$ relative to the laboratory frame.
The flow field in eq. (\ref{flow}) can be written in potential representation $\boldsymbol u(\boldsymbol r)=\boldsymbol\nabla \phi(\boldsymbol r)$, where
\begin{equation}
\label{potential}
\phi(\boldsymbol r)=-(\boldsymbol V_0\boldsymbol\cdot\boldsymbol r)\left(1+\frac{a^3}{2r^3}\right).
\end{equation}
We also have $\nabla^2\phi(\boldsymbol r)=0$ for $r>0$ due to the flow incompressibility.

We focus on the steady-state solution of Eq. (\ref{comoving}).
Multiplying eq. (\ref{comoving}) by $\exp[-\phi(\boldsymbol r)/(2D)]$, yields
\begin{equation}
\label{comoving2}
D\nabla^2 \bar c(\boldsymbol r)-\frac{u(\boldsymbol r)^2}{4D}\bar c(\boldsymbol r)+\bar A(\boldsymbol r)=0,
\end{equation}
where $\bar c(\boldsymbol r)=c(\boldsymbol r)\exp[-\phi(\boldsymbol r)/(2D)]$ and $\bar A(\boldsymbol r)=A(\boldsymbol r)\exp[-\phi(\boldsymbol r)/(2D)]$.

The original model corresponds to setting $u(\boldsymbol r)^2$ to $V_0^2$, $\phi(\boldsymbol r)$ to $-\boldsymbol V_0\boldsymbol\cdot\boldsymbol r$, and $\bar A(\boldsymbol r)$ to a point source in eq. (\ref{comoving2}).
Here we still simplify $u(\boldsymbol r)^2$ to $V_0^2$ because this term is quadratic in velocity and thus should be small close to the critical point.
We keep, however, the full expression for $\phi$ and replace the $\bar A(\boldsymbol r)$ term with a combination of a point source and a point source dipole.
The amplitude of the source dipole is chosen in a way that corresponds to an isotropic emission rate at distance $a$ from the particle center.

We thus consider the following equation
\begin{equation}
\label{comoving3}
D\nabla^2 \bar c(\boldsymbol r)-\frac{V_0^2}{4D}\bar c(\boldsymbol r)+4\pi a^2A[\delta(\boldsymbol r)+b(\boldsymbol V_0\boldsymbol\cdot\boldsymbol\nabla)\delta(\boldsymbol r)]=0.
\end{equation}
This equation can be solved analytically, yielding
\begin{equation}
\label{solution}
\bar c(\boldsymbol r)=\frac{a^2A}{Dr}\exp\left(-\frac{V_0r}{2D}\right)+b(\boldsymbol V_0\boldsymbol\cdot\boldsymbol\nabla)\left\{\frac{a^2A}{Dr}\exp\left(-\frac{V_0r}{2D}\right)\right\}
\end{equation}
The constant $b$ is found by taking the concentration field $c(\boldsymbol r)\equiv\bar c(\boldsymbol r)\exp[\phi(\boldsymbol r)/(2D)]$ and setting the first harmonic of $\boldsymbol r\boldsymbol\cdot\boldsymbol \nabla c(\boldsymbol r)$ to zero:
\begin{equation}
\label{bsolution}
b=\frac{9 a^{2}}{2D} \frac{\xi + 3}{\xi^{2} + 6 \xi + 18} \frac{(\xi-2)e^{\xi}+\xi+2}{(\xi^{2}-4\xi+8) e^{\xi} - (\xi^{2} + 4 \xi + 8)},
\end{equation}
where $\xi=3V_0a/(2D)$.
Substituting eq. (\ref{bsolution}) into eq. (\ref{solution}) yields the corrected concentration field.
We extract the first harmonic of the concentration for $r=a$ from this solution, which gives us the following expression of the swimming velocity
\begin{equation}
\label{corrected}
V_0=-\frac{18AM\left[(\xi - 2)e^{\xi} + \xi + 2\right] \exp\left(-\frac{5 \xi}{6}\right)}{D \xi^{2}\left(\xi^{2} + 6 \xi + 18\right)}=-\frac{AMaV_0}{4D^2}[1-aV_0/D+O\left((aV_0/D)^2\right)]
\end{equation}
Dividing both sides of eq. \ref{corrected} by $V_0$ and setting $V_0$ to 0 yields $-AMa/D^2\equiv Pe=4$ for the critical Peclet number, which agrees with the previous works.

\section{Finite size effect}
We consider the same phoretic model except that the size is finite.
 We  focus here only on steady state solutions in the co-moving frame with velocity $\mathbf{V}_0$. The concentration field obeys in this frame 

\begin{equation}\label{diff1p}
  D\Delta c+  \mathbf{V}_0\cdot\nabla c=-S \delta (\mathbf {r})
\end{equation} The particle is taken to move along the $z-$direction. Making the substitution $c=\bar c e^{-{zV_0\over 2D}} $ we find

\begin{equation}\label{diff1pp}
  \Delta \bar c - k^2 \bar c=-{S\over D} \delta (\mathbf {r}), \;\; k^2=\bar V_0 ^2/(4a^2) 
\end{equation}
This is the so-called screened Helmotz equation with a delta source term. The associated Green's function is defined as 
\begin{equation}\label{diff1ppp}
  \Delta G(\mathbf r,\mathbf r') - k^2 G(\mathbf r,\mathbf r')=\delta (\mathbf {r}-\mathbf r'), 
\end{equation}
We consider the domain to be finite and bounded by a sphere with radius   $r=R$ (counted from the point source). The boundary condition is taken as $\bar c (r=R)=0$. We use the eigenfunctions of the Laplacian in order to express  the Green's function.  The Laplacian eigenfunctions are spherical harmonics $Y_\ell^m(\theta, \phi)$ times spherical Bessel functions $j_\ell (r)$.  Let  $\beta_{\ell n}$ define the zero's of $j_\ell$, we have $j_\ell (\beta_{\ell n})=0$. The Laplacian eigenfunction which vanishes at $r=R$ can be written as 
 \begin{equation} \psi _{n\ell m} (r,\theta,\phi) = A _{n\ell} Y_\ell^m(\theta, \phi)   j_\ell ( \beta_{\ell n} {r/ R})  \end{equation}

Then making use of the classical method to express the Green's function in terms of eigenfunctions, we obtain
\begin{equation}\label{psiG}
 G(\mathbf r,\mathbf r')  = - \sum _{n\ell m} {2\over R^3} {1\over j^2_{\ell+1} ( \beta_{\ell m})} {Y_\ell^m(\theta, \phi)   j_\ell ( \beta_{\ell m} r/ R) Y_\ell^m(\theta', \phi')   j_\ell ( \beta_{\ell m} r'/ R) 
 \over k^2+(\beta_{\ell m} / R) ^2} \end{equation}
 Note  that the eigenvalues of the Laplacian are $(\beta_{\ell m}/ R)^2$, meaning  that the eigenvalues of the full operator  in (\ref{diff1pp}) are $ k^2+(\beta_{\ell m} / R) ^2$. The above Green's function can be rewritten as 
  \begin{equation}\label{psiGp}
 G(\mathbf r,\mathbf r')  = - {2\over R^3}\sum _{n\ell} {2\ell +1\over 4\pi} P_\ell (\cos(\gamma)) {1\over j^2_{\ell+1} ( \beta_{\ell m})} {   j_\ell ( \beta_{\ell m} r/ R)   j_\ell ( \beta_{\ell m} r'/ R) 
 \over k^2+(\beta_{\ell m} / R) ^2} \end{equation}
  after having used the addition theorem for spherical harmonics, where $P_\ell$ is the Legendre polynomial of order $\ell$ and  $\cos(\gamma )= \cos(\theta)  \cos(\theta')+  \sin(\theta)  \sin(\theta') \cos(\phi-\phi')$. Since the source term is assumed to be at the center, we set $\mathbf r'=0$, so that $ j_\ell ( \beta_{\ell m} r'/ R)   = j_\ell (0)$. Due to the properties of $j_\ell$ only $\ell=0$ survives in the sum. Using the definition of $j_0$ and $j_1$ functions, we obtain  (upon using that $\beta_{0n}= n\pi$) that the concentration field can be written as 
  
     \begin{equation}\label{ccbar}
 c(r,\theta)  = {A\over 4\pi D r} e^{-{rV_0 \cos(\theta) \over 2D} }\csch (\lvert k\rvert R) \sinh (\lvert k\rvert (R-r))
 \end{equation} 
 where we have used the result $\sum _{n=0}^{\infty} n\sin (n a)/(n^2+b^2)= \pi \csch(\pi \lvert b\rvert ) \sinh ((\pi -a)\lvert b\rvert))$, $\csch$ being the hyperbolic cosecant function. Projecting $c(r,\theta)$ on the first spherical harmonic,	and using the condition that  $ {V}_0=- M c_1/ (a\sqrt{3\pi})$ (recall that $c_1$ is the concentration contribution of the first harmonic at $r=a$) we find
  \begin{equation} \label{Vregp} {\bar V_0}=4{ {Pe}   } \left[  { \bar V_0 \cosh({\bar V_0}/2 )-2\sinh( {\bar V_0}/2) \over \bar V_0^2  }\right] \csch ( \lvert  {\bar V_0}\rvert \bar R/2) \sinh ( \lvert {\bar V_0}\rvert (\bar R-1)/2)  , \end{equation}
  where $\bar R\equiv R/a$. 
   Expanding this result  for small ${\bar V_0}$ we obtain to cubic order 
      \begin{equation}\label{vbarp}
{\bar V_0}= {Pe\over 3}{\bar V_0}  \left [ 1-\bar R ^{-1} -{{\bar V_0 }^2\over 24} (2 {\bar R} -3 +{\bar R} ^{-1} )\right ]
 \end{equation}    
 We see that the expansion is regular; the finite size has regularized the singular pitchfork behavior.
The solution  ${\bar V_0}=0$ always exists. Beyond a certain critical value $Pe=Pe_1$ there exists another solution behaving as ${\bar V_0}\sim \pm (Pe-Pe_1)^{1/2}$, with $Pe_1= 3/(1-\bar R^{-1})$. 
If  we take first the limit  $\bar R\rightarrow \infty$ in Eq. (\ref{Vregp}), we obtain 
 \begin{equation} \label{Vregpp} {\bar V_0}=-{ {Pe}  \over 2 } \left[  {\sinh( 2{\bar V_0}) \over 4 {\bar  V_0}^2} -  {\cosh(2{\bar V_0}) \over 2 {\bar V_0} }\right] e^{- 2 \lvert {\bar V_0}\rvert}  , \end{equation}
yielding  the same expression as in the main text for infinite size.
The function $\csch (\lvert \bar V_0 \rvert \bar R/2)$ has a infinite and countable set of singularities on the imaginary axis, $\bar V_0=in\pi/(2\bar R)$, $n$ being an integer.
\section{Relation between $a_k$ and $b_k$}
It is easy to obtain the general relation between $a_k$ and $b_k$. However here we only list the relations for the first three terms (generalization to arbitrary order is straightforward). The starting point is to write the Taylor expanion in terms of $a_k(\epsilon) x^{2k}$  and makes the substitution $\epsilon= x_0 (1-s)$ and $x^2=(2s-s^2) x_0^2$, so that we have
\begin{equation}
a_0[x_0(1-s)]+a_1[x_0(1-s)](2s-s^2)x_0^2+a_2[x_0(1-s)](2s-s^2)^2 x_0^4+...
\end{equation} 
Then expansing $a_k[x_0(1-s)]$ in Taylor series with respect to $s$, we obtain to leading order 
\begin{equation}
b_0(x_0)+b_1(x_0) s+b_2(x_0) s^2+...
\end{equation}
with the relations

\begin{eqnarray}
&& b_0(x_0)= a_0 ,\;\;   b_1(x_0)= a_0' +2a_1  x_0^2 ,\nonumber \\
&& b_2(x_0)= {a_0''\over 2}  -a_1 x_0^2 + 2 x_0 ^2 a_1' + 4 a_2x_0^4 
\end{eqnarray} 
where $a_k$ as well as $a_k'$ and $a_k''$, which designate first and second derivative with respect to $s$,  are evaluated at $s=0$.

 \section{2D model with consumption} 

In 2D we only need to substitute in the denominator of the propagator $\tau^{3/2}$ by $\tau$, so that the concentration field takes the form
\begin{equation}\label{diff3}
c(\mathbf {r},t)  = \int_0 ^\infty d\tau  {S \over 4\pi D\tau }\exp- { \left \{ {(\mathbf{r}+\mathbf{V}_0\tau - \mathbf{V}_0t)^2\over  4D\tau} \right\} } \; ,\end{equation}.

yielding
\begin{equation}\label{diff32}
c(\mathbf {r},t)  =  {S \over 2\pi D} K_0(\bar{\tilde r} \sqrt{\bar V_0 ^2+\epsilon^2}/4) e^{-\bar{\tilde r}\bar V_0 \cos(\theta ) /2} \; ,\end{equation}
where $\bar{\tilde r} = \tilde r/a$, and $K_0$ is the Bessel function of the second kind. Projecting  (\ref{vbarp}) on the first Fourier mode and  using the equation fixing velocity as a function of concentration (see main text) we find $ {V}_0=- 2M c_1/ (3 a)$ (where $c_1$ is the amplitude of the first Fourier mode), obtaining finally 

\begin{equation}\label{diff33}
 \bar V_0= {Pe \over 3} I_1(\bar V_0 /2) K_0(\sqrt{\bar V_0 ^2+\epsilon^2}/4)\end{equation}
where $I_1$ is the Bessel function of the first kind. Besides the trivial solution, this equation exhibits a pitchfork bifurcation. For $\epsilon=0$ the bifurcation becomes singular with $\bar V_0\sim e^{-3/Pe}$; for a small argument $I_1\sim \bar V_0$ and $K_0\sim -\ln(V_0 )$.

\section{Singularities outside of the imaginary axis}

\begin{figure}
\includegraphics[width = 0.9\columnwidth]{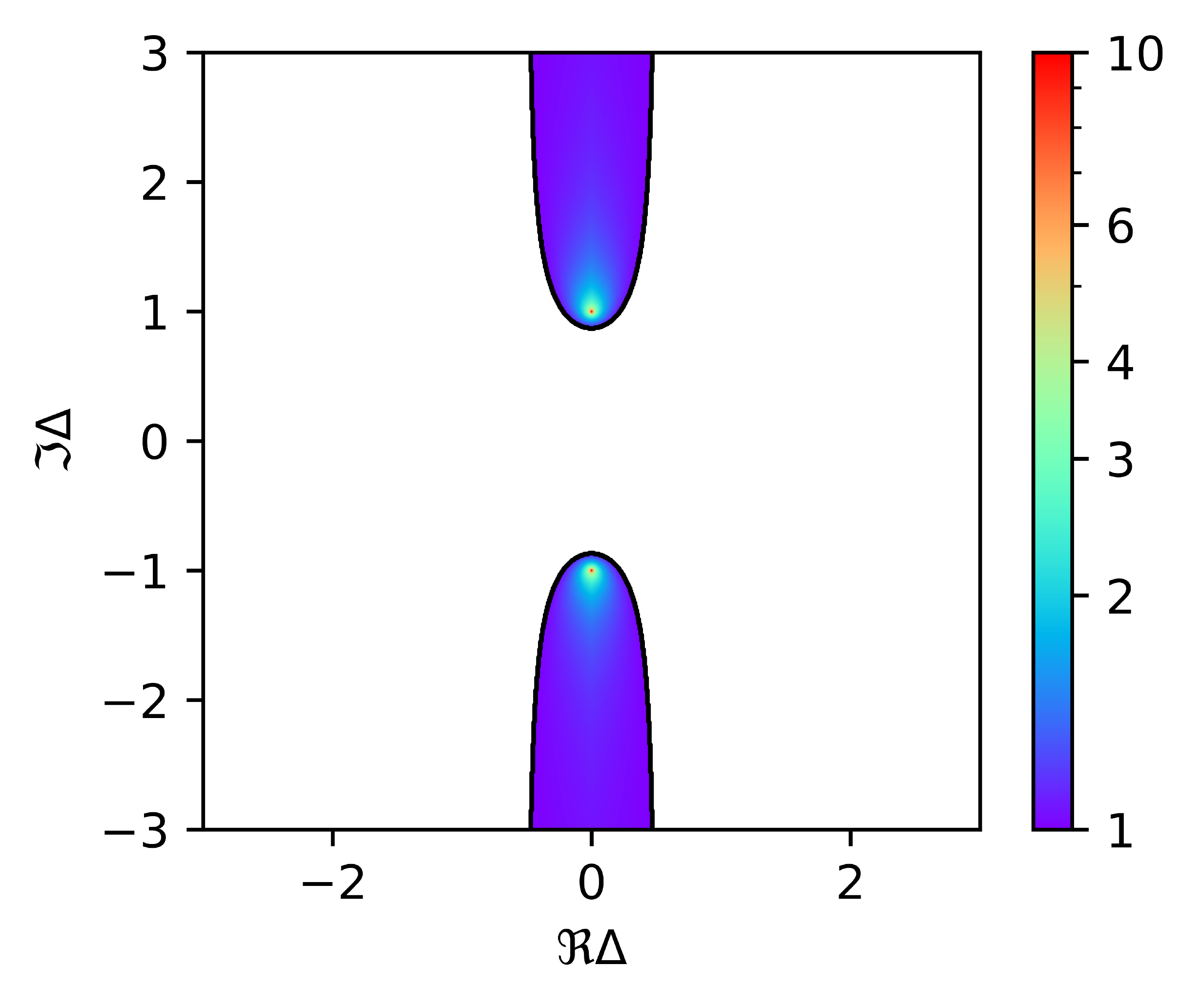}
\caption{\label{convergence}The radius of convergence of the expansion of $F$ in powers of $s$ as a function of the proportionality coefficients $\Delta_i$ setting the singular points of $F$ as a function of $\epsilon$. The black curves mark the boundary of the region of $\Delta$ for which the radius of convergence is greater than 1. Only this region is colored.}
\end{figure}

Here we discuss the applicability of the method for the problems in which the singularities are not necessary on the imaginary axis.
Suppose there is a function $F(x^2,\epsilon)$, where $x$ is the expansion parameter and $\epsilon$ is the regularization parameter, as in the Main Letter.
The function $F$ is an analytic function of $x$ with exception of singular points $x_i(\epsilon)=\Delta_i \epsilon$.
Here we allow $\Delta_i$ to be arbitrary complex numbers.
Applying the transformation $\epsilon= x_0 (1-s)$ and $x^2=(2s-s^2) x_0^2$, we obtain a function of $s$ and $x_0$.
This function is an analytical function of $s$ with exception of singular points $s_i$ given by
\begin{equation}
\label{ssingular}
s_i=1\pm\frac{1}{\sqrt{1+\Delta_i^2}}
\end{equation}
The radius of convergence of the expansion of $F$ in powers of $s$ is governed by the singular point $s_i$ with the lowest absolute value.
The success of the proposed method requires this radius of convergence to be greater than 1.
The method thus works if all $\Delta_i$ are such that $|s_i|>1$, where $s_i$ is given by (\ref{ssingular}).
Figure \ref{convergence} shows the region of the complex plane which must contain $\Delta_i$ for all singular points of $F$ in order for the expansion in $s$ to converge for $s=1$.

\end{document}